\documentclass[runningheads]{llncs}
\usepackage{cite}
\usepackage{amsmath,amssymb,amsfonts}
\usepackage{algorithmic}
\usepackage{graphicx}
\usepackage{textcomp}
\usepackage{algorithmic}
\usepackage{caption}
\usepackage{subfigure}
\usepackage{epstopdf}
\usepackage{multirow}
\usepackage{array}
\usepackage{booktabs}

\def\BibTeX{{\rm B\kern-.05em{\sc i\kern-.025em b}\kern-.08em
    T\kern-.1667em\lower.7ex\hbox{E}\kern-.125emX}}

\begin{document}

\title{Quantum Searchable Encryption for Cloud Data Based on Full-Blind Quantum Computation}
\titlerunning{}

\author{{Wenjie Liu}\inst{1,2},\and{Yinsong Xu\inst{2},\and {Wen Liu}\inst{3}, \and{Haibin Wang}\inst{1, 2} \and {Zhibin Lei}\inst{4}}}

\institute{Engineering Research Center of Digital Forensics, Ministry of Education, Nanjing 210044, P. R. China \inst{1}\\
School of Computer and Software, Nanjing University of Information Science and Technology, Nanjing 210044, P. R. China\inst{2}\\
School of Computer Science and Cybersecurity, Communication University of China, Beijing 100024, P. R. China\inst{3}\\
Hong Kong Applied Science and Technology Research Institute (ASTRI), Science Park East Avenue, Hong Kong 999077, P. R. China\inst{4}\\
Corresponding author: Wenjie Liu \email{ wenjiel@163.com}.).}

\maketitle

\begin{abstract}
Searchable encryption (SE) is a positive way to protect users sensitive data in cloud computing setting, while preserving search ability on the server side. SE allows the server to search encrypted data without leaking information about the plaintext data. In this paper, a multi-client universal circuit-based full-blind quantum computation (FBQC) model is proposed. In order to meet the requirements of multi-client accessing or computing encrypted cloud data, all clients with limited quantum ability outsource the key generation to a trusted key center and upload their encrypted data to the data center. Considering the feasibility of physical implementation , all quantum gates in the circuit are replaced with the combination of ${\pi  \mathord{\left/
 {\vphantom {\pi  8}} \right.
 \kern-\nulldelimiterspace} 8}$ rotation operator set $\{ $${R_z}({\pi  \mathord{\left/
 {\vphantom {\pi  4}} \right.
 \kern-\nulldelimiterspace} 4})$, ${R_y}({\pi  \mathord{\left/
 {\vphantom {\pi  4}} \right.
 \kern-\nulldelimiterspace} 4})$, $C{R_z}({\pi  \mathord{\left/
 {\vphantom {\pi  4}} \right.
 \kern-\nulldelimiterspace} 4})$, $C{R_y}({\pi  \mathord{\left/
 {\vphantom {\pi  4}} \right.
 \kern-\nulldelimiterspace} 4})$, $CC{R_z}({\pi  \mathord{\left/
 {\vphantom {\pi  4}} \right.
 \kern-\nulldelimiterspace} 4})$, $CC{R_y}({\pi  \mathord{\left/
 {\vphantom {\pi  4}} \right.
 \kern-\nulldelimiterspace} 4})$$\} $. In addition, the data center is only allowed to perform one ${\pi  \mathord{\left/
 {\vphantom {\pi  8}} \right.
 \kern-\nulldelimiterspace} 8}$ rotation operator each time, but does not know the structure of the circuit (i.e., quantum computation), so it can guarantee the blindness of computation. Then, through combining this multi-client FBQC model and Grover searching algorithm, we continue to propose a quantum searchable encryption scheme for cloud data. It solves the problem of multi-client access mode under searchable encryption in the cloud environment, and has the ability to resist against some quantum attacks. To better demonstrate our scheme, an example of our scheme to search on encrypted 2-qubit state is given in detail. Furthermore, the security of our scheme is analysed from two aspects: external attacks and internal attacks, and the result indicates that it can resist against such kinds of attacks and also guarantee the blindness of data and computation.
\end{abstract}


\section{Introduction}
\label{sec:introduction}
In recent years, cloud computing has achieved great development in different fields, such as wireless networks \cite{Wang19,Xu19Li,Li19}, IoT \cite{Ota18,Qi16,Xu19Liu,Ren18,Xu19DAD}, resource allocation \cite{Xu18He,Qi19Chen,Xu19Zhang} and so on. As it provides economic and convenient service, more and more clients are planning to upload their data onto the public clouds now. And with the popularity of mobile devices, more and more companies can push the content they want based on the data uploaded by clients, which greatly promotes the development of mobile Internet \cite{Qi19Wang,Gong18,Xu19SSP,Qi18}. However, data stored in the cloud server may suffer from malicious use by cloud service providers since data owners have no longer direct control over data. For instance, data stored in a bank must not be arbitrarily obtained \cite{Wang19A,Fu17,Wang19B}. Considering data privacy and security, it is a recommended practice for data owners to encrypt data before uploading onto the cloud \cite{Bosch14}. Therefore, an efficient search technique for encrypted data is extremely urgent.

A popular way to search over encrypted data is searchable encryption (SE), which is desirable to support the fullest possible search functionality on the server side, without decrypting the data, and thus, with the smallest possible loss of data confidentiality. The first searchable encryption was proposed by Song \textit{et al.} \cite{Song00}. This scheme uses stream ciphers and pseudo-random functions to implement ciphertext retrieval, but it also has a series of problems, such as low search efficiency and data privacy. Therefore, Goh \cite{Goh03} built a index structure based on the Bloom filter to achieve fast retrieval of ciphertext data. However, the Bloom filter itself has a certain error rate, and the result returned by the cloud server to the data user may not be accurate. Besides, Curtmola \textit{et al.} \cite{Curtmola06} and Boneh \textit{et al.} \cite{Boneh04} use the idea of "keyword-file" to construct a symmetric searchable encryption scheme and a public key search able encryption scheme, respectively. Both schemes have significant improvements in safety and efficiency. Nowadays, many researchers have tried to use kNN algorithm \cite{Xia16,Xu19E}, user interest model \cite{Fu16}, blockchain technology \cite{Chen19,Wang19Guo}, multi-keyword ranked search \cite{Cao14} and so on, to improve the search efficiency and data privacy.

However, with the development of quantum computation, the powerful computing power of quantum computers poses an increasingly strong threat to public key systems \cite{Shor97} and symmetric key systems \cite{Kaplan16,Dong19}. Besides, quantum computation are also applied in other fields, such as quantum key agreement (QKA) \cite{Huang17,Liu18Xu}, quantum steganography (QS) \cite{Wu18,Qu18Zhu}, quantum machine learning \cite{Liu18Gao,Liu19Gao}, and so on. Especially, to protect the privacy of client's data, many researchers have proposed a novel quantum computation model: blind quantum computation (BQC), where the client with limited quantum resources can perform quantum computation by delegating the computation to an untrusted quantum server, and the privacy of the client can still be guaranteed. BQC can be generally divided into two categories: one is the measurement-based blind quantum computation (MBQC), and the other is the circuit-based blind quantum computation (CBQC). In MBQC, measurement is the main driving force of computation, which follows the principle of "entangle-measure-correct", and a certain number of quantum qubits are entangled to form a standard graph state \cite{Broadbent09,Kong16,Kashefi17}.

Different from MBQC, CBQC is based on the quantum circuit, which is composed of many kinds of quantum gates \cite{Arrighi06,Tan17,Fisher14,Broadbent15,Zhang18,Liu18,Sheng18,Fitzsimons17,Li18}. Among them, Fisher \cite{Fisher14} and Broadbent \cite{Broadbent15} firstly proposed a representative CBQC model: delegating quantum computation (DQC). In their schemes, an untrusted server can perform arbitrary quantum computations on encrypted quantum bits (qubits) without learning any information about the inputs, where the quantum computations are implemented by a universal set of quantum gates ($X$, $Z$, $H$, $S$, $T$, \textit{CNOT}). And then the client can easily decrypt the results of the computation with the decryption key. Then, Tan \textit{et al.}\cite{Tan17} give 3 circuits of other quantum gates (\textit{CZ}, \textit{SWAP}, and \textit{Toffoli}) for blind quantum computation. However, in their schemes, the server knows the content of delegating computation. To further protect computation privacy, a few universal circuit-based "full-blind" quantum computation (FBQC) schemes are proposed \cite{Zhang18,Liu18}, i.e., the server also does not know the content of delegated computation. These two schemes use two different strategies to achieve full blindness. Zhang \textit{et al.}'s scheme decomposes all quantum gates into several basic rotation operators, and inserts trap qubits and trap gates to achieve full blindness, where the trap gate is composed of basic rotation operators and does not affect the computation results. In Liu \textit{et al.}'s scheme, the client uses the strategy of oblivious mechanism to make the computation blind, where the desirable delegated quantum operation, one of $\{ H,P,CNOT,T\} $, is replaced by a fixed sequence ($H$, $P$, \textit{CZ}, \textit{CNOT}, $T$). However, all these mentioned CBQC schemes are only a single-client model, i.e., clients can only delegate the server to compute their own data, which are not convenient for different clients to compute others' data.

In order to implement multi-client universal circuit-based FBQC for searchable encryption, i.e., different clients can store or search their data in the quantum cloud server, we propose a quantum searchable encryption scheme for cloud data based on full-blind quantum computation (QSE-FBQC). Clients with limited quantum ability firstly use $X$ and $Z$ gates to encrypt their data with the encryption keys generated by the key center, and then upload the encrypted data to the data center. The data center performs search computation on the encrypted data if other clients need, where the search computation are implemented by a universal set of quantum gates ($X$, $Z$, $H$, $S$, $T$, \textit{CNOT}, \textit{CZ}, \textit{Toffoli}). But, the data center only performs one ${\pi  \mathord{\left/
 {\vphantom {\pi  8}} \right.
 \kern-\nulldelimiterspace} 8}$ rotation operator from ${\pi  \mathord{\left/
 {\vphantom {\pi  8}} \right.
 \kern-\nulldelimiterspace} 8}$ rotation operator set $\{ $${R_z}({\pi  \mathord{\left/
 {\vphantom {\pi  4}} \right.
 \kern-\nulldelimiterspace} 4})$, ${R_y}({\pi  \mathord{\left/
 {\vphantom {\pi  4}} \right.
 \kern-\nulldelimiterspace} 4})$, $C{R_z}({\pi  \mathord{\left/
 {\vphantom {\pi  4}} \right.
 \kern-\nulldelimiterspace} 4})$, $C{R_y}({\pi  \mathord{\left/
 {\vphantom {\pi  4}} \right.
 \kern-\nulldelimiterspace} 4})$, $CC{R_z}({\pi  \mathord{\left/
 {\vphantom {\pi  4}} \right.
 \kern-\nulldelimiterspace} 4})$, $CC{R_y}({\pi  \mathord{\left/
 {\vphantom {\pi  4}} \right.
 \kern-\nulldelimiterspace} 4})$$\} $ on qubits sent by the key center each time, and sends these qubits back to the key center. Repeating this process multiple times can complete any quantum gate in the circuit of search computation. This kind of strategy can make the data center unable to know the positions and the orders of quantum gates in the circuit, which guarantees the blindness of computation. When the search computation finishes, the key center generate corresponding decryption key. Finally, the clients who need the search result from the data center, also use $X$ and $Z$ gates to decrypt the encrypted search result with the decryption key.

The rest of the paper is organized as follows. Sect. \ref{sec2} provides some preliminary knowledge about quantum computation and circuit-based blind quantum computation. Then, a quantum searchable encryption scheme for cloud data based on full-blind quantum computation is proposed in Sect. \ref{sec3}. Moreover, we give a concrete example that use Grover algorithm to search on encrypted 2-qubit state in Sect. \ref{sec4}. And the security of our scheme is analysed in Sect. \ref{sec5}. Sect. \ref{sec6} is devoted to compare our scheme with some existing SE schemes and BQC schemes. Finally, Sect. \ref{sec7} gives discussion and conclusion of this paper.

\section{Preliminaries}
\label{sec2}
\subsection{Quantum computation}
In quantum compution, the quantum bit (called qubit) \cite{Nielsen02} is the basic unit of quantum information and has two possible states $\left| 0 \right\rangle $ and $\left| 1 \right\rangle $, which is often referred to as quantum superposition state,

\begin{equation}\label{eqn1}
  \left| \varphi  \right\rangle  = \alpha \left| 0 \right\rangle  + \beta \left| 1 \right\rangle,
\end{equation}
where $\alpha $, $\beta $ are complex numbers, and ${\left| \alpha  \right|^2} + {\left| \beta  \right|^2} = 1$. $\left| 0 \right\rangle $ and $\left| 1 \right\rangle $ can be represented by vectors,

\begin{equation}
  \left| 0 \right\rangle  = \left[ \begin{array}{l}
1\\
0
\end{array} \right],  {\kern 1pt} {\kern 1pt} {\kern 1pt} {\kern 1pt} {\kern 1pt} {\kern 1pt} {\kern 1pt} {\kern 1pt} {\kern 1pt} {\kern 1pt} {\kern 1pt} {\kern 1pt} {\kern 1pt} {\kern 1pt} {\kern 1pt} \left| 1 \right\rangle  = \left[ \begin{array}{l}
0\\
1
\end{array} \right].
\label{eqn2}
\end{equation}
Then, $\left| \varphi  \right\rangle$ can be expressed in vector form $\left| \varphi  \right\rangle  = \left( \begin{smallmatrix}
\alpha \\
\beta
\end{smallmatrix} \right)$.

With the information carrier (qubit), we also need some quantum gates to implement the information processing. For single-qubit gates, we have \textit{Pauli-X}, \textit{Pauli-Z}, \textit{H} (\textit{Hadamard}), \textit{S} and \textit{T} gates, which can be described as $2 \times 2$ unitary matrices as below,

\begin{equation}
\begin{split}
&\textit{X} = \left[ {\begin{matrix}
0 & 1  \\
1 & 0
\end{matrix}} \right],{\kern 4pt} \textit{Z} = \left[ {\begin{matrix}
 1&0 \\
  0&-1 \\
\end{matrix}} \right],{\kern 4pt}
\textit{H}  =  \frac{1}{{\sqrt 2 }}\left[ {\begin{matrix}
1&1\\
1&{ - 1}
\end{matrix}} \right],{\kern 4pt}\\
&\textit{S}  =  \left[ {\begin{matrix}
1&0\\
0&{i}
\end{matrix}} \right],{\kern 4pt}
\textit{T}  =  \left[ {\begin{matrix}
1&0\\
0&{{e^{{{i\pi } \mathord{\left/
 {\vphantom {{i\pi } 4}} \right.
 \kern-\nulldelimiterspace} 4}}}}
\end{matrix}} \right].
\label{eqn3}
\end{split}
\end{equation}

Especially, Ref. \cite{Nielsen02} also points out that for arbitrary unitary operator $U$ performed on single-qubit, there exist $\theta $, $\alpha $, $\beta $ and $\gamma $, s.t.

\begin{equation}
\begin{split}
U &= {e^{i\theta }}{R_z}(\alpha ){R_y}(\beta ){R_z}(\gamma )\\
& = \left( {\begin{matrix}
   {{e^{i(\theta  - \frac{\alpha }{2} - \frac{\gamma }{2})}}\cos \frac{\beta }{2}} & { - {e^{i(\theta  - \frac{\alpha }{2} + \frac{\gamma }{2})}}\sin \frac{\beta }{2}}  \\
   {{e^{i(\theta  + \frac{\alpha }{2} - \frac{\gamma }{2})}}\sin \frac{\beta }{2}} & {{e^{i(\theta  + \frac{\alpha }{2} + \frac{\gamma }{2})}}\cos \frac{\beta }{2}}  \\
\end{matrix}} \right)
\label{eqn4}
\end{split}
\end{equation}

\begin{figure}[htbp]
  \centering
  \includegraphics[width=3in]{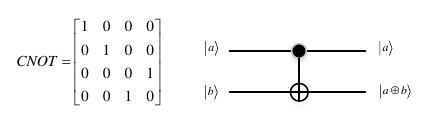}\\
  \caption{Matrix representation and quantum circuit of \textit{CNOT} gate.}\label{fig1}
\end{figure}

\begin{figure}[htbp]
  \centering
  \includegraphics[width=3in]{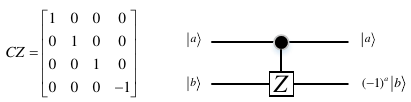}\\
  \caption{Matrix representation and quantum circuit of \textit{CZ} gate.}\label{fig2}
\end{figure}

For double-qubit gates, the commonly used multi-qubit gates are \textit{CNOT} and \textit{CZ} gates. The matrix representations and quantum circuits of \textit{CNOT} and \textit{CZ} are shown in Fig. \ref{fig1} and Fig. \ref{fig2}, respectively.

\begin{figure}[htbp]
  \centering
  \includegraphics[width=3in]{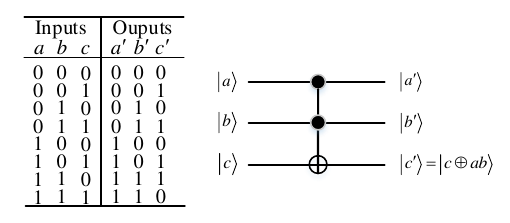}\\
  \caption{Truth table and quantum circuit of \textit{Toffoli} gate.}\label{fig3}
\end{figure}

Finally, for triple-qubit gates, \textit{Toffoli} gate is another frequently used multi-qubit gate, which is illustrated in Fig. \ref{fig3}. With these single-qubit gates and multi-qubit gates, we can implement arbitrary quantum computation.

\subsection{Circuit-based blind quantum computation}
\begin{figure}[htbp]
  \centering
  \includegraphics[width=3in]{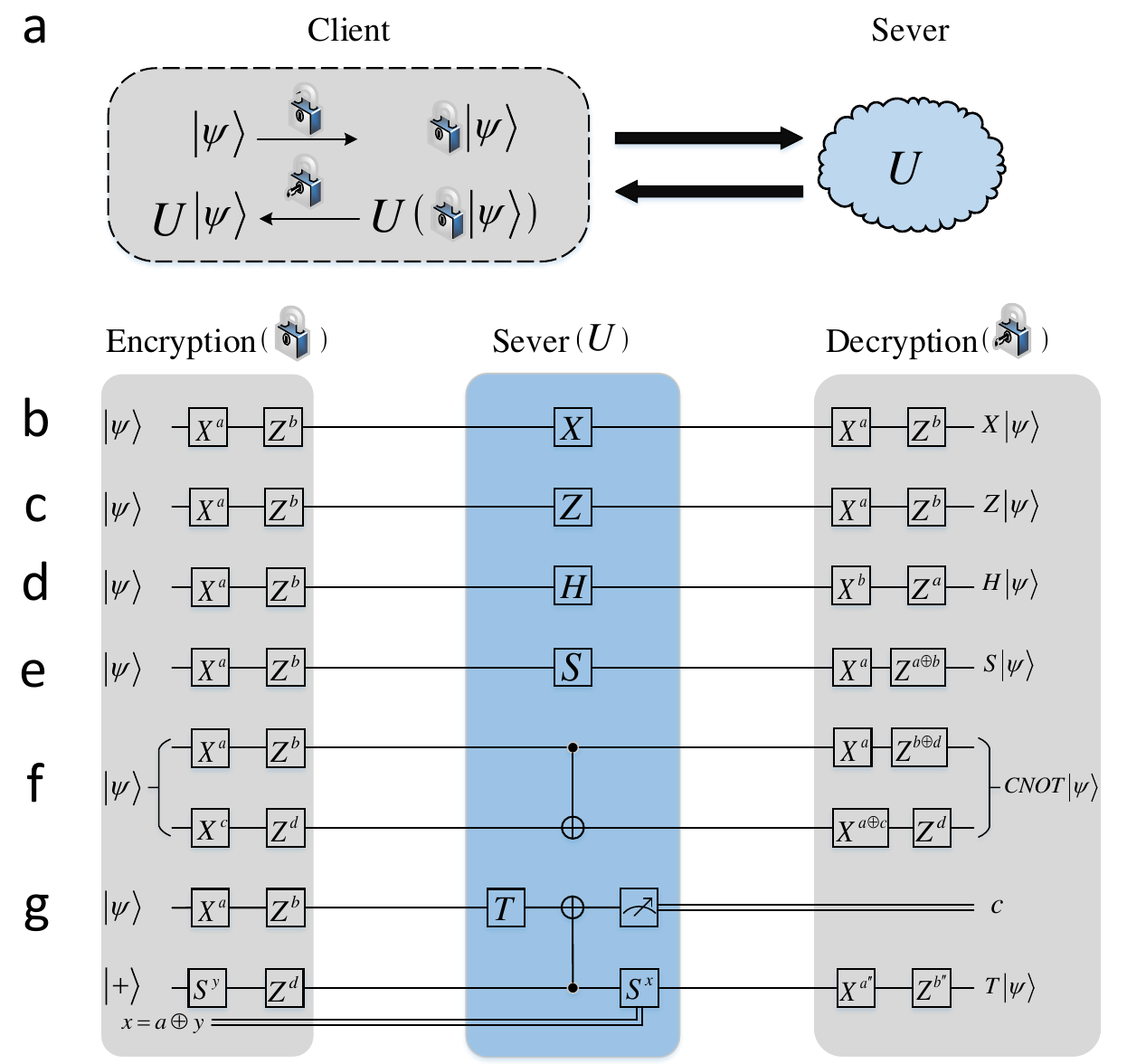}\\
  \caption{The process of blind quantum computation for each quantum gate in Fisher's and Broadbent's schemes}\label{fig4}
\end{figure}

Fisher \cite{Fisher14} and Broadbent \cite{Broadbent15} proposed a specific blind quantum computation scheme based on quantum circuit. It (see Fig. \ref{fig4} a) starts with a client who has quantum information that needs to be sent to a remote server for processing. The client first encrypts one input qubit $\left| \psi  \right\rangle $ and sends it to a quantum server, who performs a computation $U$ on the encrypted qubit. The server returns the state which the client decrypts to get $U\left| \psi  \right\rangle $.

In the scheme, to encrypt a qubit $\left| \psi  \right\rangle $, a client applies a combination of Pauli $X$ and $Z$ operations to get a encrypted qubit ${X^a}{Z^b}\left| \psi  \right\rangle $, where $a,b \in \{ 0,1\} $ (as well as $c,d \in \{ 0,1\} $ for the \textit{CNOT} gate in Fig. \ref{fig4}f). Then, the server perform quantum computing $U$, which is composed of unitary operations from the Clifford group $\{ X,Z,H,S,\textit{CNOT}\} $ and one additional non-Clifford gate, $T$ gate. As shown in Fig. \ref{fig4} b-f, when $U \in \{ X,Z,H,S,\textit{CNOT}\} $, clifford gates do not require any additional resources, and decryption is straightforward. However, when $U = T$ (see Fig. \ref{fig4} g), the server requires the client to send an auxiliary qubit ${Z^d}{P^y}\left|  +  \right\rangle $, where $y,d \in \{ 0,1\} $. to control a \textit{CNOT} gate with the encrypted qubit. The server measures the encrypted qubit and outcome $c \in \{ 0,1\} $ is returned to the client, which is used in decryption. The client sends a single classical bit, $x = a \oplus y$, to control a $S$ gate on the auxiliary qubit, which is returned to the client as ${X^{a''}}{Z^{b''}}R\left| \psi  \right\rangle $, where $a'' = a \oplus c$ and $b'' = a(c \oplus y \oplus 1) \oplus b \oplus d \oplus y$.

\begin{figure}[htbp]
  \centering
  \includegraphics[width=3.2in]{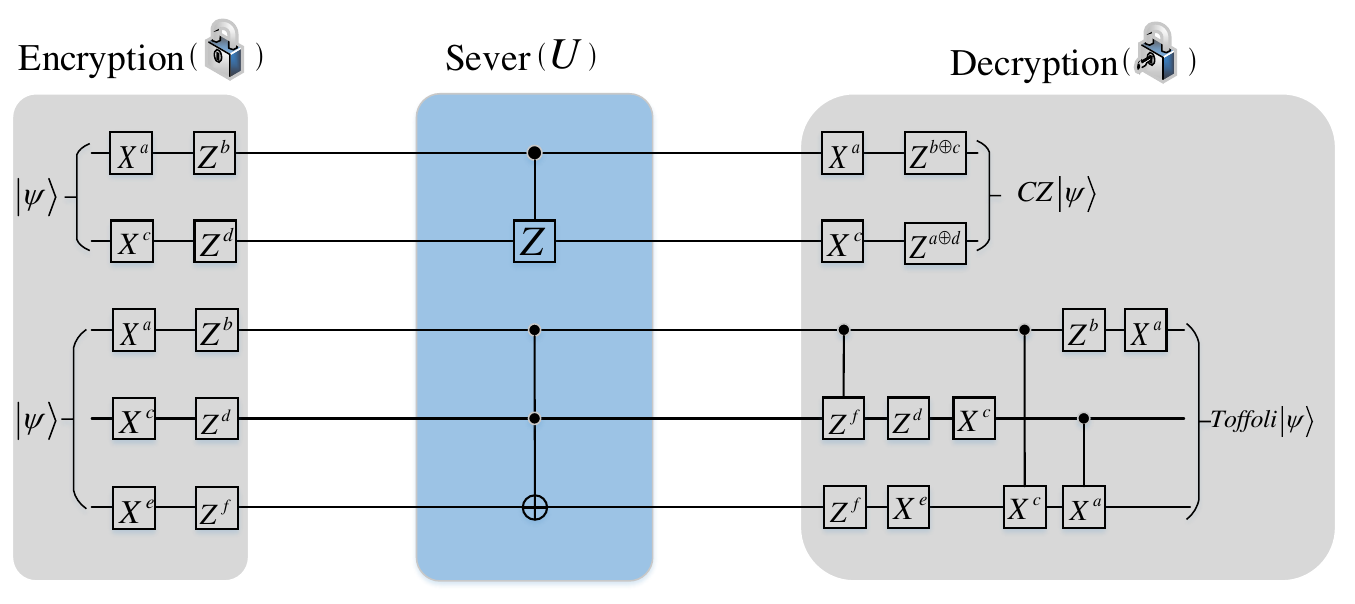}\\
  \caption{The process of blind quantum computation for \textit{CZ} and \textit{Toffoli} gate}\label{fig5}
\end{figure}

In Ref.\cite{Tan17}, Tan \textit{et al.} give the circuit of \textit{CZ} and \textit{Toffoli} gates for blind quantum computation,which is illustrated in Fig. \ref{fig5}.

\section{Quantum searchable encryption for cloud data based on universal full-blind quantum computation}
\label{sec3}
\subsection{Multi-client universal circuit-based full-blind quantum computation model}
\begin{figure}[htbp]
  \centering
  \includegraphics[width=3.2in]{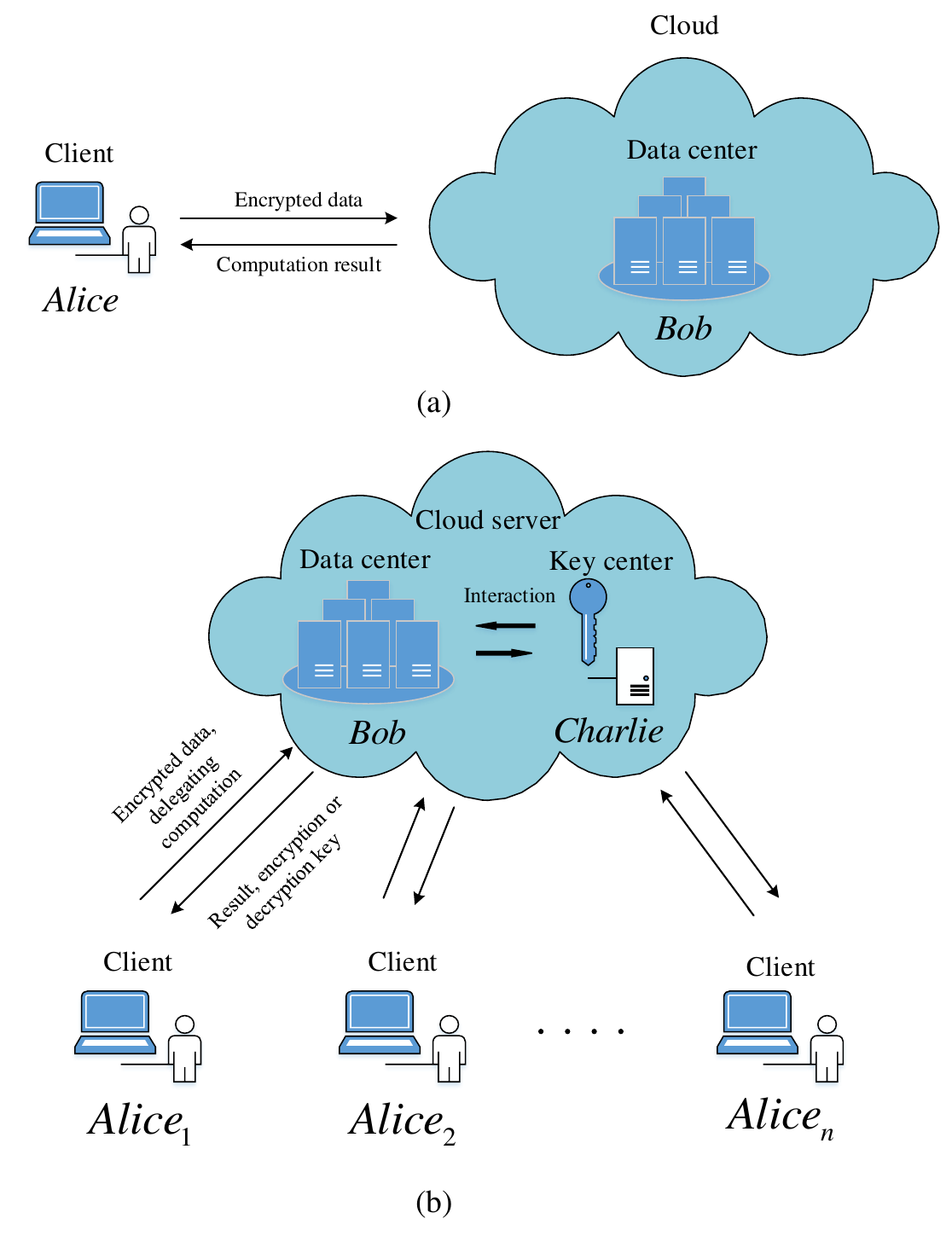}\\
  \caption{Single-client model (a) and multi-client model (b) of circuit-based BQC schemes}\label{fig6}
\end{figure}

\begin{figure*}[htbp]
  \centering
  \includegraphics[width=6in]{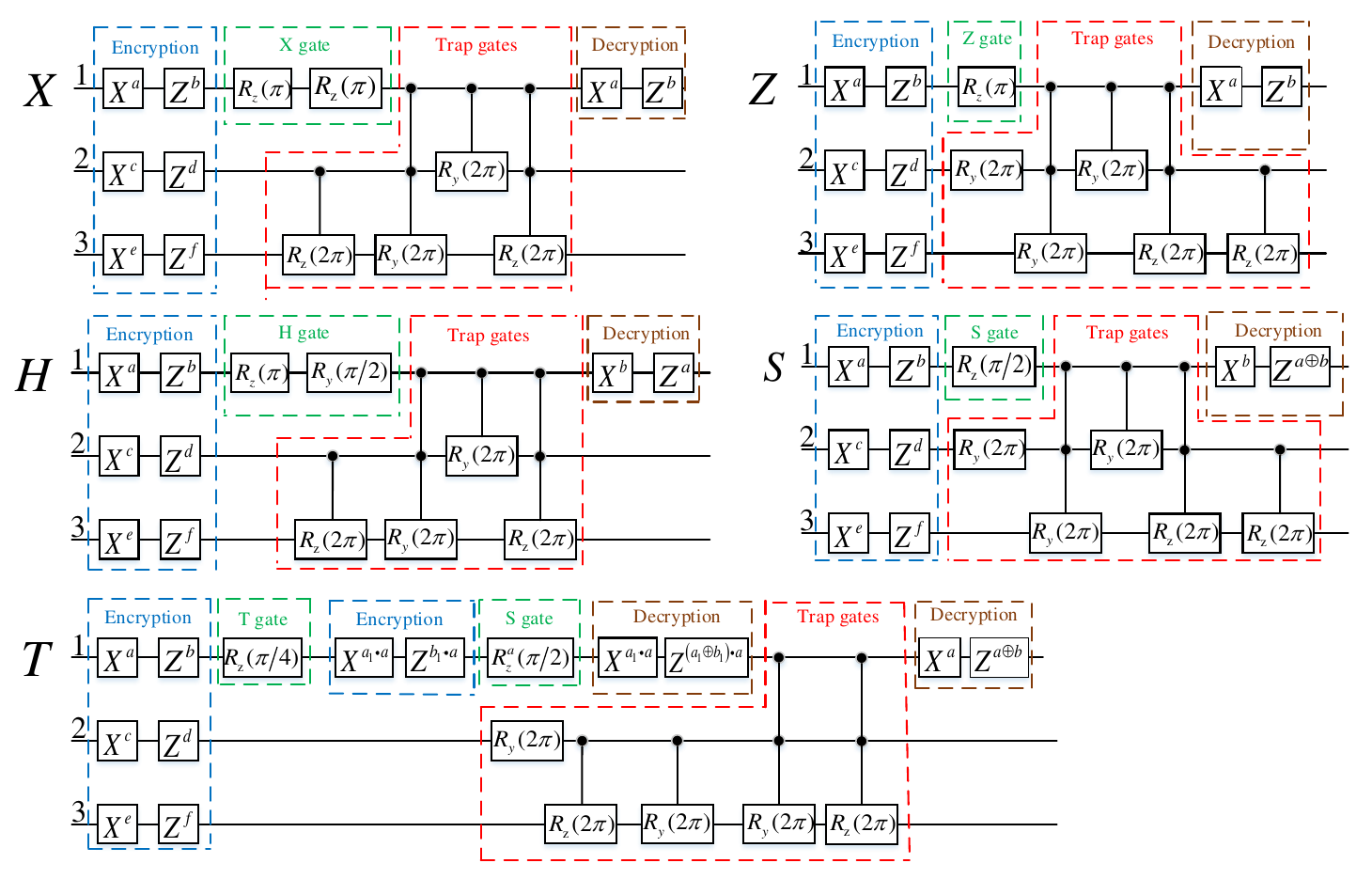}\\
  \caption{Our model for single-qubit gates $X$, $Z$, $H$, $S$ and $T$ ($a$, $b$, $c$, $d$, $e$, $f$ $ \in $ $\{ 0,1\} $), where the sign 1 belongs to $\left| \psi  \right\rangle $ and the signs 2 and 3 belong to trap qubits, respectively. And the blue, green, red and brown dotted line indicates the encryption operations, the actual operations, the trap operations and the decryption operations, respectively.}\label{fig7}
\end{figure*}

As mentioned above, these circuit-based BQC schemes (including circuit-based FBQC schemes) are only a single-client model as shown in Fig. \ref{fig6} (a), i.e., clients only delegate server to compute over their own encrypted data in this model, which are not convenient to data sharing. Besides, clients need to interact with server and compute decryption key frequently, which consume a large amount of computing and communication resources. To solve these two problems and keep full-blind, we propose a multi-client universal circuit-based full-blind quantum computation model (see Fig. \ref{fig7} (b)), where clients ($Alic{e_1}$, $Alic{e_2}$, $ \cdots $, $Alic{e_n}$) outsource key generation to a trusted key center ($Charlie$) and let key center interact with the server (i.e., data center $Bob$). $Charlie$ and $Bob$ compose the quantum cloud servers.

Suppose a client $Alic{e_1}$ as the data owner uploads her encrypted data $\left| \psi  \right\rangle $ to $Bob$, and another client $Alic{e_2}$ delegates $Bob$ to perform quantum computation on $\left| \psi  \right\rangle $ to get the computation result. The basic process of our scheme is given as below.

Step 1: $Alic{e_1}$ sends a number $n$ (the number of qubits in $\left| \psi  \right\rangle $) to $Charlie$.

Step 2: $Charlie$ sends back a string of $2n$ random binary bits back to $Alic{e_1}$ by BB84 protocol \cite{Bennett84}. The $2n$ bits of the binary string act as $ek = ({x_i},{z_i})$, where ${x_i},{z_i} \in \{ 0,1\} $ and $i \in \{ 1,2, \cdots ,n\} $.

Step 3: $Alic{e_1}$ encrypts her data $\left| \psi  \right\rangle $ with $ek$ and sends encrypted state ${E_{ek}}\left| \psi  \right\rangle  =  \otimes _{i = 1}^nX_i^{{x_i}}Z_i^{{z_i}}\left| \psi  \right\rangle $ to $Bob$.

Step 4: $Alic{e_2}$ asks for $Charlie$ to delegate $Bob$ to perform quantum computation over ${E_{ek}}\left| \psi  \right\rangle $, and $Charlie$ gets ${E_{ek}}\left| \psi  \right\rangle $ from $Bob$.

\begin{figure*}[htbp]
  \centering
  \includegraphics[width=4.5in]{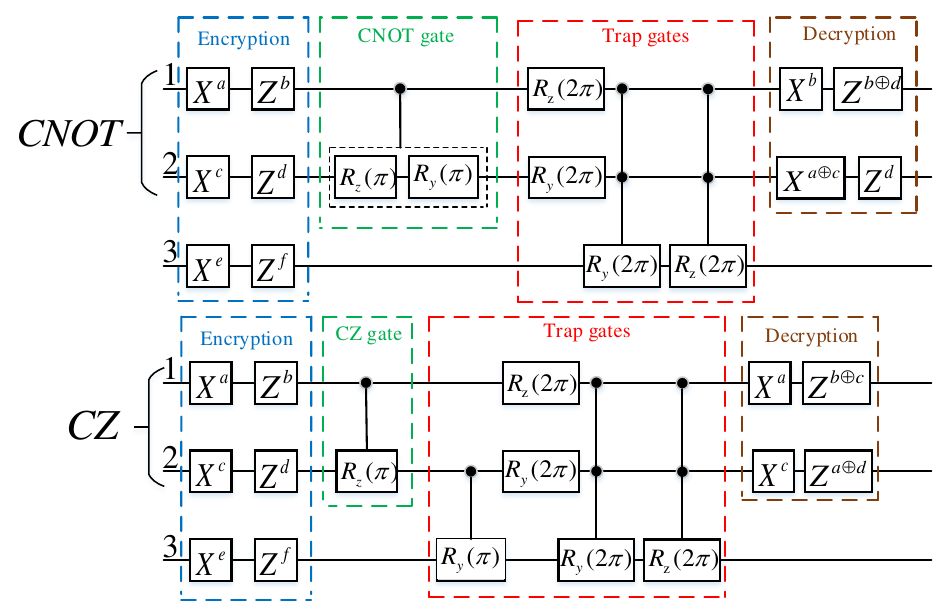}\\
  \caption{Our model for double-qubit gates \textit{CZ} and \textit{CNOT}.}\label{fig8}
\end{figure*}

\begin{figure*}[htbp]
  \centering
  \includegraphics[width=6in]{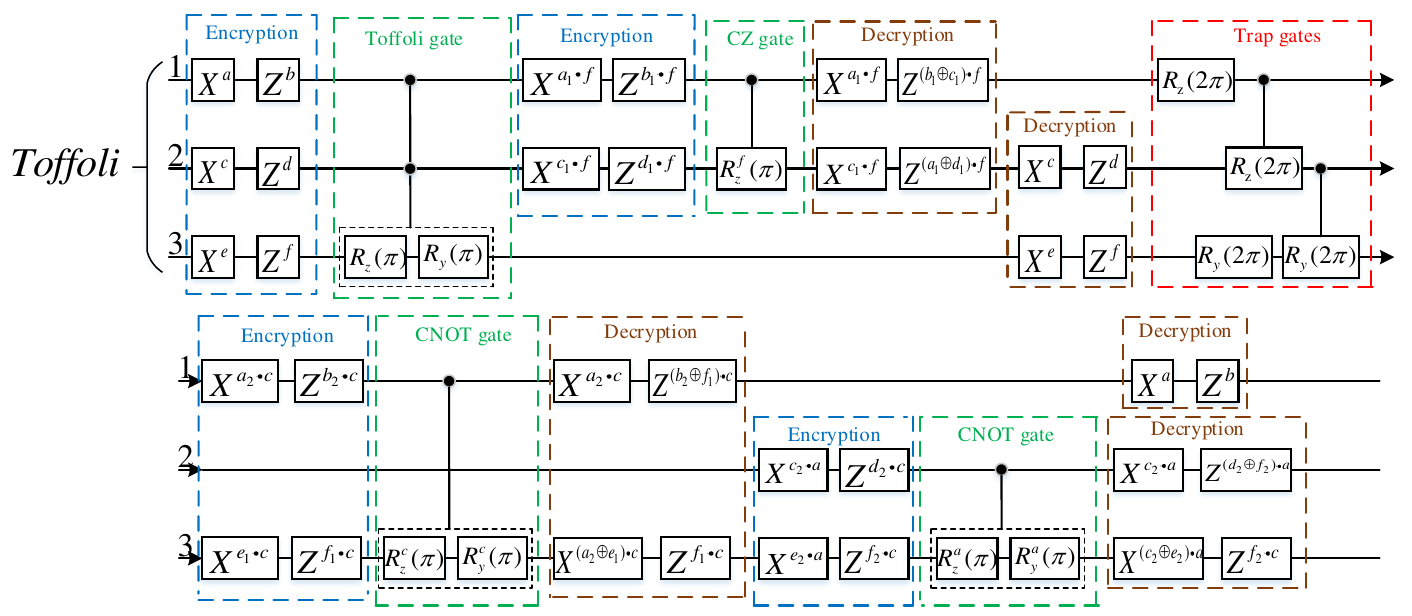}\\
  \caption{Our model for triple-qubit gate \textit{Toffoli}. }\label{fig9}
\end{figure*}

Step 5: If $Charlie$ wants $Bob$ to perform ${G_i}$ (or ${G_{i,l}}$, or ${G_{i,l,k}}$) gate on ${\left| \psi  \right\rangle _i}$ (or ${\left| \psi  \right\rangle _{i,l}}$, or ${\left| \psi  \right\rangle _{i,l,k}}$), $Charlie$ needs to insert two trap qubits (or one trap qubits, or no trap qubits) into ${\left| \psi  \right\rangle _i}$ (or ${\left| \psi  \right\rangle _{i,l}}$, or ${\left| \psi  \right\rangle _{i,l,k}}$) randomly (see Fig. \ref{fig7}, \ref{fig8} and \ref{fig9}), where $G$ $ \in $ $\{ $$X$, $Z$, $H$, $S$, $T$, \textit{CNOT}, \textit{CZ}, \textit{Toffoli}$\} $, $i$, $l$ and $k$ represents the $i$th, $l$th and $k$th qubit in $\left| \psi  \right\rangle $ respectively.

Step 6: $Charlie$ sends 3 prepared qubits to $Bob$ and instructs $Bob$ to perform rotation operator ${U_i}$ (or ${U_{i,l}}$, or ${U_{i,l,k}}$) on these 3 qubits, where $U$ $ \in $ $\{ $${R_z}({\pi  \mathord{\left/
 {\vphantom {\pi  4}} \right.
 \kern-\nulldelimiterspace} 4})$, ${R_y}({\pi  \mathord{\left/
 {\vphantom {\pi  4}} \right.
 \kern-\nulldelimiterspace} 4})$, $C{R_z}({\pi  \mathord{\left/
 {\vphantom {\pi  4}} \right.
 \kern-\nulldelimiterspace} 4})$, $C{R_y}({\pi  \mathord{\left/
 {\vphantom {\pi  4}} \right.
 \kern-\nulldelimiterspace} 4})$, $CC{R_z}({\pi  \mathord{\left/
 {\vphantom {\pi  4}} \right.
 \kern-\nulldelimiterspace} 4})$, $CC{R_y}({\pi  \mathord{\left/
 {\vphantom {\pi  4}} \right.
 \kern-\nulldelimiterspace} 4})$$\} $, where ${R_z}({\mu _1}{\pi  \mathord{\left/
 {\vphantom {\pi  4}} \right.
 \kern-\nulldelimiterspace} 4}) = {R_z}{({\pi  \mathord{\left/
 {\vphantom {\pi  4}} \right.
 \kern-\nulldelimiterspace} 4})^{{\mu _1}}}$ and ${R_y}({\mu _2}{\pi  \mathord{\left/
 {\vphantom {\pi  4}} \right.
 \kern-\nulldelimiterspace} 4}) = {R_y}{({\pi  \mathord{\left/
 {\vphantom {\pi  4}} \right.
 \kern-\nulldelimiterspace} 4})^{{\mu _2}}}$, $i$, $l$ and $k$ represents the $i$th, $l$th and $k$th qubit in these 3 qubits, respectively.

Step 7: $Bob$ performs rotation operator $U$ on 3 qubits, and sends back to $Charlie$.

Step 8: $Charlie$ performs $X$ and $Z$ operations on 3 qubits as needed.

Step 9: Repeat Step 6 to 8 until $Charlie$ gets $( \otimes _{i = 1}^nX_i^{{{x'}_i}}Z_i^{{{z'}_i}}) \otimes G\left| \psi  \right\rangle $ and generates decryption key $dk = ({{x'}_i},{{z'}_i})$, where $G$ gate can be composed of a plurality of $U$ gates (as shown in Eq. \ref{eqn5}) and ${{x'}_i},{{z'}_i} \in \{ 0,1\} $. When $U$ is used as a trap gate, it can be executed in any order.
\begin{equation}
\begin{split}
&X = {e^{\frac{{i\pi }}{2}}}{R_y}(\pi ){R_z}(\pi ), Z= {e^{\frac{{i\pi }}{2}}}{R_z}(\pi ){R_y}(0),\\
&H = {e^{\frac{{i\pi }}{2}}}{R_y}(\frac{\pi }{2}){R_z}(\pi ), S = {e^{\frac{{i\pi }}{4}}}{R_z}(\frac{\pi }{2}){R_y}(0),\\
&T = {e^{\frac{{i\pi }}{8}}}{R_z}(\frac{\pi }{4}){R_y}(0),\\
&C{Z_{12}} = {[C - ({e^{\frac{{i\pi }}{2}}}{R_z}(\pi ){R_y}(0))]_{12}},\\
&CNO{T_{12}} = {[C - ({e^{\frac{{i\pi }}{4}}}{R_z}(\frac{\pi }{2}){R_y}(0))]_{12}},\\
&Toffol{i_{12,3}} = {[CC - X]_{12,3}},
\label{eqn5}
\end{split}
\end{equation}
where ${R_y}(0) = {R_z}(0) = I$, $ - {R_y}(2\pi ) =  - {R_y}(\frac{\pi }{4}) = I$, ${R_y}(\pi ) = {R_y}{(\frac{\pi }{4})^4}$, ${R_y}(\frac{\pi }{2}) = {R_y}{(\frac{\pi }{4})^2}$, ${R_z}(\frac{\pi }{2}) = {R_z}{(\frac{\pi }{4})^2}$, ${R_z}(\pi ) = {R_z}{(\frac{\pi }{4})^4}$, ${(C - {R_y}(\theta ))_{12}}$= $\left( \begin{smallmatrix}
1&0&0&0 \\
0&1&0&0\\
0&0&\cos \frac{\theta }{2}&\sin \frac{\theta }{2}\\
0&0&\sin \frac{\theta }{2}&\cos \frac{\theta }{2}
\end{smallmatrix} \right)$ and ${(C - {R_z}(\theta ))_{12}}$= $\left( \begin{smallmatrix}
1&0&0&0 \\
0&1&0&0\\
0&0&{e^{ - \frac{{i\theta }}{2}}}&0\\
0&0&0&{e^{\frac{{i\theta }}{2}}}
\end{smallmatrix} \right)$. In our scheme, the global phase factors (${e^{\frac{{i\theta }}{2}}}$) are ignored.

Step 10: Repeat Step 5 to 9 until $Charlie$ gets ${E_{dk}}\left| {result} \right\rangle  = ( \otimes _{i = 1}^nX_i^{{{x'}_i}}Z_i^{{{z'}_i}})( \otimes _{j = 1}^s{G_j})\left| \psi  \right\rangle  = ( \otimes _{i = 1}^nX_i^{{{x'}_i}}Z_i^{{{z'}_i}})\left| {result} \right\rangle $ and generates final decryption key $dk = ({{x'}_i},{{z'}_i})$, where $s$ represents the number of $G$ gate in the circuit of quantum computation, $j \in \{ 1,2, \cdots ,s\} $ and $\left| {result} \right\rangle $ is the computation result which $Alic{e_2}$ needs.

Step 11: $Charlie$ sends ${E_{dk}}\left| {result} \right\rangle $ and $dk$ to $Alic{e_2}$, where $dk$ is transformed by BB84.

Step 12: $Alic{e_2}$ decrypts the encrypted result ${E_{dk}}\left| {result} \right\rangle $ with $dk$, to get $\left| {result} \right\rangle $.

To better understand our multi-client universal FBQC scheme, we give three examples of delegating $Bob$ to perform one single-qubit gate (\textit{X} gate in Fig. \ref{fig7}), one double-qubit gate (\textit{CZ} gate in Fig. \ref{fig8}), and one triple-qubit gate (\textit{Toffoli} gate in Fig. \ref{fig9}), respectively. For the sake of simplicity, we mainly explain Step 5 to 9 in these examples.

\subsubsection{single-qubit gate - \textit{X gate}}
1. $Charlie$ sends 3 encrypted qubits $X_1^aZ_1^b{\left| \psi  \right\rangle _1} \otimes X_2^cZ_2^d{\left| \phi  \right\rangle _2} \otimes X_3^eZ_3^f{\left| \phi  \right\rangle _3}$ to $Bob$ (${\left| \phi  \right\rangle _2}$ and ${\left| \phi  \right\rangle _3}$ are trap qubits, and their positions are arbitrary.).

\noindent2. $Bob$ performs four rounds (i.e., Step 6 to 8) ${R_z}{(\frac{\pi }{4})_1}$ on qubit $X_1^aZ_1^b{\left| \psi  \right\rangle _1}$ and returns 3 qubits to himself through $Charlie$.

\noindent3. $Bob$ performs four rounds ${R_y}{(\frac{\pi }{4})_1}$ on ${R_z}{(\pi )_1}{X^a}{Z^b}{\left| \psi  \right\rangle _1}$ and returns 3 qubits to himself through $Charlie$.

\noindent4. To confuse the actual delegating operations, $Bob$ also performs eight rounds ${(C - {R_z}(\frac{\pi }{4}))_{23}}$, ${(CC - {R_y}(\frac{\pi }{4}))_{12,3}}$, ${(C - {R_y}(\frac{\pi }{4}))_{12}}$ and ${(CC - {R_z}(\frac{\pi }{4}))_{12,3}}$ on 3 qubits. Note, all trap gates are randomly inserted without affecting the original circuits.

\noindent5. $Charlie$ gets $(X_1^aZ_1^b){X_1}{\left| \psi  \right\rangle _1}$ from $Bob$ and generates decryption key $dk = {(a,b)_1}$.

\subsubsection{double-qubit gate - \textit{CZ gate}}
1. $Charlie$ sends 3 encrypted qubits $(X_1^aZ_1^b \otimes X_2^cZ_2^d){\left| \psi  \right\rangle _{1,2}} \otimes X_3^eZ_3^f{\left| \phi  \right\rangle _3}$ to $Bob$ (${\left| \phi  \right\rangle _3}$ is a trap qubit.).

\noindent2. Bob performs four rounds ${(C - {R_z}(\frac{\pi }{4}))_{12}}$ on $(X_1^aZ_1^b \otimes X_2^cZ_2^d){\left| \psi  \right\rangle _{1,2}}$ and returns 3 qubits to himself through $Charlie$.

\noindent3. To confuse the actual delegating operations, $Bob$ also performs eight rounds ${(C - {R_z}(\frac{\pi }{4}))_{23}}$, ${(CC - {R_y}(\frac{\pi }{4}))_{12,3}}$, ${(C - {R_y}(\frac{\pi }{4}))_{12}}$ and ${(CC - {R_z}(\frac{\pi }{4}))_{12,3}}$ on 3 qubits.

\noindent4. $Charlie$ gets $(X_1^aZ_1^{b \oplus c} \otimes X_2^cZ_2^{a \oplus d})C{Z_{12}}{\left| \psi  \right\rangle _{1,2}}$ from $Bob$ and generates decryption key $dk = \{ {(a,b \oplus c)_1},{(c,a \oplus d)_2}\} $.

\subsubsection{triple-qubit gate - \textit{Toffoli gate}}
1. $Charlie$ sends encryption quantum states $(X_1^aZ_1^b \otimes X_2^cZ_2^d \otimes X_3^eZ_3^f){\left| \psi  \right\rangle _{1,2,3}}$ to $Bob$.

\noindent2. $Bob$ performs four rounds ${(CC - {R_z}(\frac{\pi }{4}))_{12,3}}$ on $(X_1^aZ_1^b \otimes X_2^cZ_2^d \otimes X_3^eZ_3^f){\left| \psi  \right\rangle _{1,2,3}}$ and returns 3 qubits to himself through $Charlie$.

\noindent3. $Bob$ performs four rounds ${(CC - {R_y}(\frac{\pi }{4}))_{12,3}}$ and returns 3 qubits to himself through $Charlie$.

\noindent4. When $f = 1$, $Charlie$ should encrypt 3 qubits again by performing $(X_1^{{a_1} \cdot f}Z_1^{{b_1} \cdot f}) \otimes (X_2^{{c_1} \cdot f}Z_2^{{d_1} \cdot f})$ on first 2 qubits and send them to $Bob$. It is the same for the similar case (when $a = c = 1$). $Bob$ performs four rounds ${(C - R_z^f(\frac{\pi }{4}))_{12}}$ and returns 3 qubits to $Charlie$. $Charlie$ needs to decrypt by performing $(X_1^{{a_1} \cdot f}Z_1^{({b_1} \oplus {c_1}) \cdot f}) \otimes (X_2^{{c_1} \cdot f}Z_2^{({a_1} \oplus {d_1}) \cdot f})$ on first 2 qubits.

\noindent4. $Charlie$ performs ${I_1} \otimes X_2^cZ_2^d \otimes X_3^eZ_3^f$ on 3 qubits and returns them to $Bob$.

\noindent5. $Bob$ performs four rounds ${(C - R_z^c(\frac{\pi }{4}))_{13}}$ and ${(C - R_y^c(\frac{\pi }{4}))_{13}}$, and returns to himself through $Charlie$.

\noindent6. Bob performs four rounds ${(C - R_z^a(\frac{\pi }{4}))_{23}}$ and ${(C - R_y^a(\frac{\pi }{4}))_{23}}$, and returns to himself through $Charlie$.

\noindent7. To confuse the actual delegating operations, $Bob$ performs eight rounds ${R_z}{(2\pi )_1}$, ${R_y}{(2\pi )_3}$, ${(C - {R_z}(\frac{\pi }{4}))_{12}}$ and  ${(C - {R_y}(\frac{\pi }{4}))_{23}}$.

\noindent8. $Charlie$ gets $(X_1^aZ_1^b) \otimes Toffol{i_{12,3}}{\left| \psi  \right\rangle _{1,2,3}}$ from $Bob$ and generates decryption key $dk = \{ {(a,b)_1},{(0,0)_2},{(0,0)_3}\} $.

As can be seen from the above, through trap gates, trap qubits and ways of constant interaction, $Bob$ cannot distinguish these qubits from $Charlie$ every time whether belong to the same original quantum state. Meanwhile, he also cannot distinguish which rotation operators belong to the same original quantum gate.

\subsection{Quantum searchable encryption for cloud data based on full-blind quantum computation}
We have established a multi-client universal circuit-based FBQC framework for easy data sharing. To achieve the aim of searchable encryption, we propose a concrete quantum searchable encryption scheme for cloud data based on full-blind quantum computation (QSE-FBQC). For the sake of simplicity, we take four roles (the data owner $Alic{e_1}$, the data searcher $Alic{e_2}$, the data center $Bob$ and the key center $Charlie$) as an example to describe our scheme. The specific process of QSE-FBQC scheme is as follows and shown in Fig. \ref{fig10}.

1. $Alic{e_1}$ sends a number $n$ to $Charlie$, where $n$ is the number of qubits which she wants to encrypt.

2. $Charlie$ sends back a string of $2n$ random binary bits back to $Alic{e_1}$ by BB84 \cite{Bennett84}. The $2n$ bits of the binary string act as $ek = ({x_i},{z_i})$, where ${x_i},{z_i} \in \{ 0,1\} $ and $i \in \{ 1,2, \cdots ,n\} $.

3. $Alic{e_1}$ encrypts her plain data $\left| \psi  \right\rangle  = \frac{1}{{\sqrt M }}\sum\limits_{j = 0}^{M - 1} {\left| {j,data(j)} \right\rangle } $ with $ek$ and sends encrypted state ${E_{ek}}\left| \psi  \right\rangle  = \frac{1}{{\sqrt M }}({I^{ \otimes m}} \otimes ( \otimes _{i = 1}^nX_i^{{x_i}}Z_i^{{z_i}}))\sum\limits_{j = 0}^{M - 1} {\left| {j,data(j)} \right\rangle } $ to $Bob$, where the item index $j$ within $\left| \psi  \right\rangle $ is not encrypted and composed of $m$ qubits, $M = {2^m}$, ${data(j)}$ is the data and composed of $n$ qubits.

\begin{figure}[htbp]
  \centering
  \includegraphics[width=3.2in]{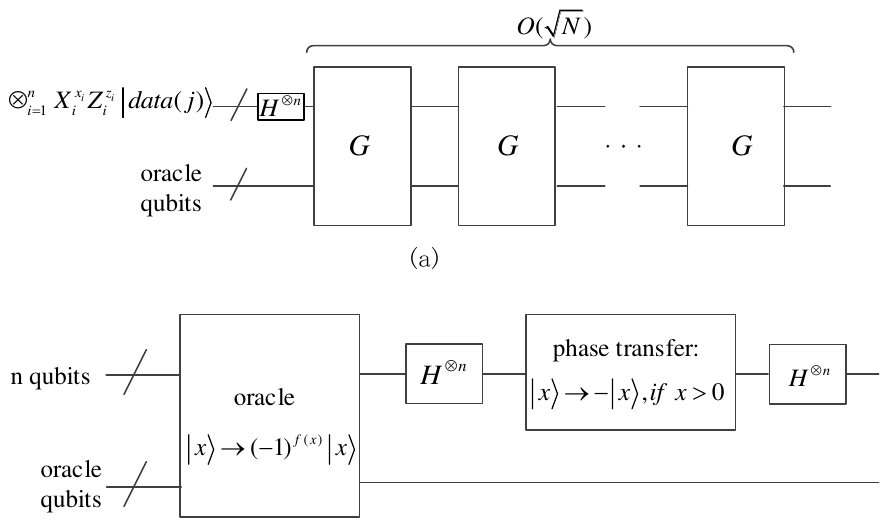}\\
  \caption{Schematic circuit for Grover algorithm. (b) is the schematic circuit for $G$ in (a). }\label{fig11}
\end{figure}

\begin{figure}[htbp]
  \centering
  \includegraphics[width=3.2in]{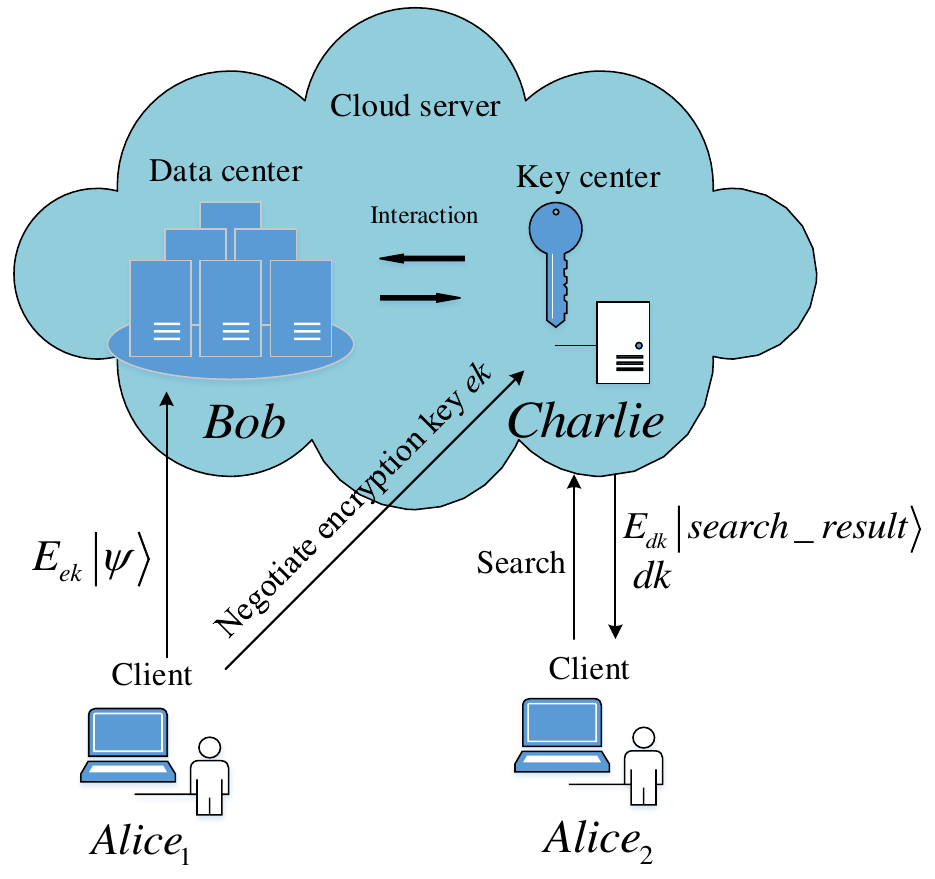}\\
  \caption{The process of our QSE-FBQC scheme}\label{fig10}
\end{figure}

4. $Alic{e_2}$ wants $Bob$ to search over ${E_{ek}}\left| \psi  \right\rangle $, and $Charlie$ interacts with $Bob$ according to the quantum circuit of search computing, which is as same as Step 5 to 10. The search computation can be composed of Grover algorithm, which is illustrated in Fig. \ref{fig11}. For a search space of $N = {2^n}$ elements and one solution, we only need to apply the search oracle $O(\sqrt N )$ times to obtain a solution. During the interaction between Bob and Charlie, the sender needs to add decoy qubits to the data and record their location. When the receiver receives the data, the sender announces the states and locations of the decoy qubits (selected from $\left\{ {\left| 0 \right\rangle ,\left| 1 \right\rangle ,\left|  +  \right\rangle ,\left|  -  \right\rangle } \right\}$), and the receiver confirms whether it is the same as the published state by measuring the state of the decoy qubits. If they are the same, the receiver proceeds to the next step; otherwise, the sender resends the data.

5. When the search computation is completed, $Charlie$ sends the search result ${E_{dk}}\left| {search\_result} \right\rangle $ and decryption key $dk$ to $Alic{e_2}$, where $dk$ is transformed by BB84.

6. $Alic{e_2}$ uses $dk$ to decrypt the state ${E_{dk}}\left| {search\_result} \right\rangle $, and gets $\left| {search\_result} \right\rangle $.

\section{An example of QSE-FBQC scheme}
\label{sec4}
\begin{figure*}[htbp]
  \centering
  \includegraphics[width=6in]{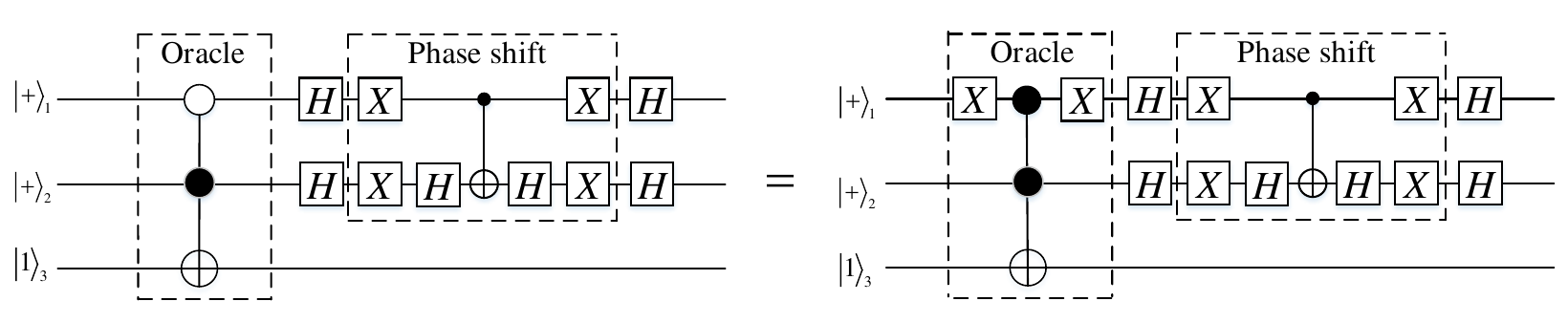}\\
  \caption{The circuit of Grover algorithm for ${\left| 0 \right\rangle _1}{\left| 1 \right\rangle _2}$ from ${\left|  +  \right\rangle _1}{\left|  +  \right\rangle _2}$.}\label{fig12}
\end{figure*}

Suppose $Alic{e_1}$ has a set {00, 01, 10, 11} and $Alic{e_2}$ wants to find item 01 from this set. It is equivalent to using Grover algorithm to find $\left| 0 \right\rangle \left| 1 \right\rangle $ from $\left|  +  \right\rangle \left|  +  \right\rangle $, and the circuit of search computation is shown in Fig. \ref{fig12}. The example proceeds in seven steps provided below.

1. $Alic{e_1}$ sends a number 2 to $Charlie$.

2. $Charlie$ sends a string of 4 random binary bits (act as $ek$) back to $Alic{e_1}$ by BB84 protocol.

3. $Alic{e_1}$ encrypts her data $\left| \psi  \right\rangle  = {\left|  +  \right\rangle _1}{\left|  +  \right\rangle _2}$ with $X$ and $Z$ gates according to $ek$.

\begin{figure*}[htbp]
  \centering
  \includegraphics[width=6in]{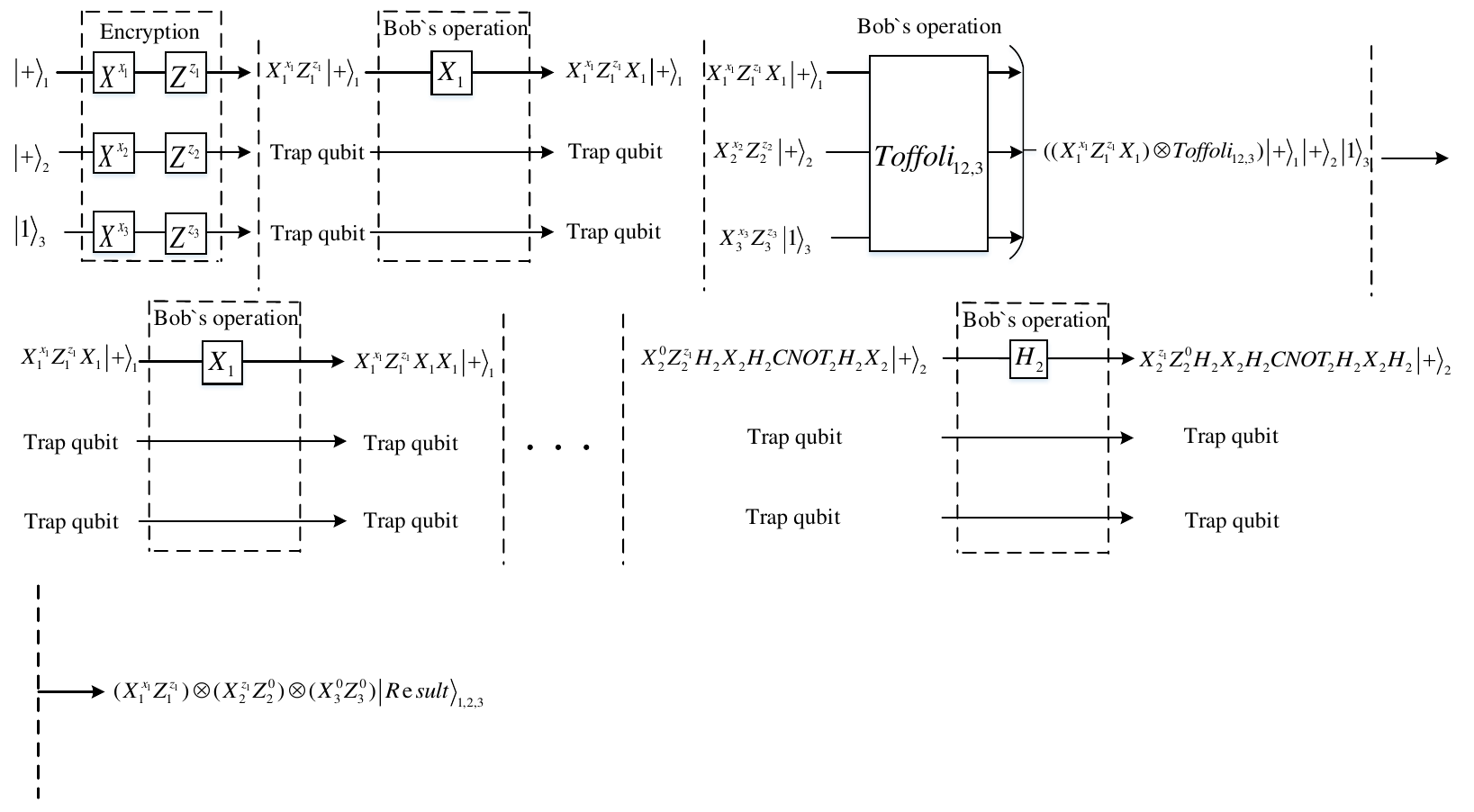}\\
  \caption{The whole process of search computing on ${\left|  +  \right\rangle _1}{\left|  +  \right\rangle _2}{\left| 1 \right\rangle _3}$.}\label{fig13}
\end{figure*}

4. $Alic{e_2}$ wants to get ${\left| 0 \right\rangle _1}{\left| 1 \right\rangle _2}$ from ${E_{ek}}\left| \psi  \right\rangle $, and $Charlie$ interacts with $Bob$ according to the quantum circuit of search computing in Fig. \ref{fig12}. $Charlie$ needs to insert an auxiliary qubit $\left| 1 \right\rangle $ into $\left| \psi  \right\rangle $ firstly. The whole process of search computing is shown in Fig. \ref{fig13}.

5. When the search is completed, $Charlie$ sends the search result state $(X_1^{{x_1}}Z_1^{{z_1}}) \otimes (X_2^{{z_1}}Z_2^0) \otimes (X_3^0Z_3^0){\left| {result} \right\rangle _{1,2,3}}$ and $dk = \{ {({x_1},{z_1})_1},{({z_1},0)_2},{(0,0)_3}\} $ to $Alic{e_2}$.

6. $Alic{e_2}$ uses $dk$ to decrypt the state $(X_1^{{x_1}}Z_1^{{z_1}}) \otimes (X_2^{{z_1}}Z_2^0) \otimes (X_3^0Z_3^0){\left| {result} \right\rangle _{1,2,3}}$, and abandons the auxiliary third qubit to get ${\left| 0 \right\rangle _1}{\left| 1 \right\rangle _2}$ (the global phase factors are ignored).

\section{Security analysis}
\label{sec5}
In this section, the security of the proposed QSE-FBQC scheme is analyzed as below. We analyze the security from two aspects: external attacks and internal attacks. The former refers to the attacks by the eavesdroppers outside the protocol, while the latter refers to the attacks by the data center in the scheme.

\subsection{External attack}
Let Eve be an eavesdropper, who tries to get some information about clients' data. To get the information, he firstly needs to get encryption or decryption key from $Charlie$, because all of the transmitted data are encrypted by the key. First, he can perform the intercept-and-resend attack by intercepting all qubits sent from $Charlie$ and resending fake qubits to the client in Step 2 or 11. However, due to the use of BB84 protocol, all qubits are encoded into $X$ or $Z$ basis according to the classical key, Eve cannot distinguish which basis each qubit belongs to and gets nothing form these qubits. And the client can check for the existence of such attack by measuring the received fake qubits. Then, the client can abandon this key and ask $Charlie$ to regenerate a new key.

Besides, Eve may also perform the intercept-and-resend attack during the communication between $Bob$ and $Charlie$. Due to the existence of tarp qubits and trap gates, Eve cannot distinguish which qubits are not trap qubits, and which required operations are actual operations. But this kind of attack will destroy the results of the delegated computation because of fake qubits sent by Eve. $Charlie$ can insert decoy qubits randomly, which consist of $X$-basis and $Z$-basis, and record the positions of them. Similar to BB84 protocol, when $Bob$ gets all qubits, $Charlie$ announce the positions and the state of decoy qubits. $Bob$ can check for the existence of such attack with a higher probability by measuring the decoy qubits. For example, if the number of decoy qubits used for eavesdropping checking is $m$, the success probability of detecting the existence of Eve is $1 - {\left( {{3 \mathord{\left/
 {\vphantom {3 4}} \right.
 \kern-\nulldelimiterspace} 4}} \right)^m}$, which obviously increases with the increase of $m$, and which is close to 1 when $m$ is large enough.

\subsection{Internal attack}
The internal attack mainly caused by $Bob$, who wants to know the information about the data and the computation. In a sense, to immunize internal attacks is actually to ensure the blindness of the data and the computation, which is analysed as below.

\subsubsection{The blindness of data}
For the blindness of data in our scheme, $Alice$ performs encryption operations $X$ and $Z$ on $n$-qubit state $\left| \psi  \right\rangle $, and then sends these encrypted qubits $ \otimes _{i = 1}^nX_i^{{x_i}}Z_i^{{z_i}}\left| \psi  \right\rangle $ to $Bob$. Although $Bob$ intercepts them, he does not know the value of $({x_i},{z_i})$ ($i \in \{ 1,2, \cdots ,n\} $), he still cannot get anything from the encrypted data.

However, the circuit of $T$ gate and \textit{Toffoli} gate for blind quantum computation is special. Because $Charlie$ is not able to perform the $S$, \textit{CNOT} and \textit{CZ} corrections, respectively, which the three operations should be delegated to $Bob$. Once $Bob$ obtains the information of corrections, then the encryption keys of encrypted qubits are exposed. So, $Charlie$ needs to encrypt qubits again with $X$ and $Z$ operations when the $S$, \textit{CNOT} and \textit{CZ} corrections need to be delegated. Therefore, $Bob$ can not distinguish whether these qubits belong to the original quantum state and get nothing from the encrypted qubits.

\subsubsection{The blindness of computation}
The computation that the client wants to implement can be seen as a desirable circuit which is made up of the delegated quantum gates, therefore the blindness of computation is equivalent to the blindness of the delegated quantum gates. In order to make the delegated quantum gates blind, these quantum gates ($G$ gate) can be decomposed into the combination of rotation operators ($U$ gate). $Bob$ performs partial rotation operators in every round, which can compose the actual $G$ gate or trap gates, so he does not know what is the correct gate. That is, $Charlie$ can successfully hide quantum computation process.

Without loss of generality, we give an simple example to explain our model. Suppose $Charlie$ wants to delegate quantum gates $H$, $X$, \textit{CZ} and \textit{Toffoli} to $Bob$, while $Bob$ cannot know the data and the content of computation in our model. The data have already been all encrypted by $Alice$ with gates $X$ and $Z$. Let all rotation operators have labels according to the order in every quantum circuit from left to right. In quantum circuit of gate $H$, the performing order of these rotation operators is ${h_1}$, ${h_2}$, $ \cdots $, ${h_m}$. In quantum circuit of gate $X$, the performing order of these rotation operators is ${x_1}$, ${x_2}$, $ \cdots $, ${x_n}$. In quantum circuit of gate \textit{CZ}, the performing order of these rotation operators is ${c_1}$, ${c_2}$, $ \cdots $, ${c_d}$. In quantum circuit of gate \textit{Toffoli}, the performing order of these rotation operators is ${t_1}$, ${t_2}$, $ \cdots $, ${t_w}$. Note that, ${h_i}$, ${x_i}$, ${c_i}$ and ${t_i}$ $ \in $ $U$. In the model, all gates are started in an arbitrary way, i.e., the process is randomly designed by $Charlie$, such as ${t_1}$, ${x_1}$, ${c_1}$, ${h_1}$, ${c_2}$, ${x_2}$, $ \cdots $, ${t_w}$, ${x_n}$, ${c_d}$, ${h_m}$. $Bob$ cannot distinguish these qubits from $Charlie$ every time whether belong to the same original quantum states. Meanwhile, he also cannot distinguish which rotation operators belong to a quantum gate. Therefore, $Bob$ cannot know what gates are realized in our model.

\section{Performance evaluation}
\label{sec6}
In order to evaluate our scheme, we chose two classical searchable encryption (SE) schemes \cite{Cao14,Fu16} and two blind quantum computation (BQC) schemes \cite{Broadbent15,Liu18} as references, and compare our QSE-FBQC scheme with them from the following aspects: time complexity of index construction, time complexity of search , full-blind, multi-client access and eavesdropping detection.

The classical SE schemes generally consists of three main parts: index construction, trapdoor generation and search. Since the process of index construction and trapdoor generation are to encrypt the keyword set $W$ extracted from data or query keywords ${\tilde W}$ through the encryption key $SK$, the time complexity of each process is similar. Therefore, we only consider the aspect of the time complexity of index construction. Suppose that the number of entries for the data is $N$. In Cao \textit{et al.}'s and Fu \textit{et al.}'s SE scheme, the major computation in the phase of index construction includes the splitting procedure and two multiplications of a $\left( {cN + u + 1} \right) \times \left( {cN + u + 1} \right)$ matrix and a $\left( {cN + u + 1} \right)$ vector, where $c$ and $u$ are constants, and ${cN}$ represents the number of keywords in $W$. So the time complexity is $O({N^2})$. However, it does not need the process of index construction and trapdoor generation, and make search computation over encrypted data directly in Broadbent's BQC scheme, Liu \textit{et al.}'s BQC scheme and our scheme. So the time complexity of index construction is 0. On the other hand, although Fu \textit{et al.}'s SE scheme based on user interest model is more efficient than Cao \textit{et al.}'s scheme when users request more relevant data, the time complexity of search are both $O(N)$. But, these mentioned BQC schemes and our scheme use Grover algorithm to make a quadratic speedup in search, so the time complexity are $O(\sqrt N )$. For a more intuitive representation, the results of the comparison are shown in Table \ref{tab1}.

As we can see, BQC schemes and our scheme are significantly more efficient than the classical SE schemes in the aspect of search efficiency. However, these BQC schemes do not support multi-client access, which is not convenient for data sharing in cloud environment. And all clients in our scheme outsource the key generation to a trusted key center, which is easy to make each client get search result by the decryption key from the key center. Besides, although Liu \textit{et al.} consider that the desirable delegated quantum operation, one of $\{ H,P,CNOT,T\} $, is replaced by a fixed sequence $(H,P,CZ,CNOT,T)$ to make the computation blind, they do not consider detecting eavesdroppers when two parties communicate with each other in their scheme. Our scheme takes the strategy of inserting decoy qubits into transmitted data to check for eavesdropping behavior.

\begin{table*}[htbp]
\centering
\caption{Comparison with classical SE schemes and BQC schemes }\label{tab1}
\setlength{\tabcolsep}{1mm}
\begin{tabular}{cccccc}
\hline
\multirow{2}{*}{Schemes} & \multirow{2}{*}{Cao \textit{et al.}'s}& \multirow{2}{*}{Fu \textit{et al.}'s}& \multirow{2}{*}{Broadbent's}& \multirow{2}{*}{Liu \textit{et al.}'s}& \multirow{2}{*}{Our scheme} \\
\specialrule{0em}{1pt}{1pt}
&SE scheme\cite{Cao14}&SE scheme\cite{Fu16} &BQC scheme\cite{Broadbent15} &BQC scheme\cite{Liu18}& \\
\specialrule{0em}{1pt}{1pt}
\hline
Time complexity of index construction  &  $O({N^2})$ &$O({N^2})$ &0 & 0& 0\\
Time complexity of search              &  $O(N)$ &$O(N)$ &$O(\sqrt N )$& $O(\sqrt N )$& $O(\sqrt N )$\\
Full-blind quantum computation         & No  &No &No&Yes& Yes\\
Multi-client access                    & Yes  &Yes &No&No& Yes\\
Eavesdropping detection                & No &No &No&No& Yes\\
\hline
\end{tabular}
\end{table*}

\section{Discussion and conclusion}
\label{sec7}
In this paper, we firstly propose a multi-client circuit-based full-blind quantum computation model, and then apply this model on the searchable encryption to get a QSE-FBQC scheme. In our scheme, different clients with limited quantum ability can upload their encrypted data to a powerful but untrusted quantum data center and the data center can search on the encrypted data without decryption. Besides, the data center also cannot know what search computation he has implemented by himself, i.e., making the computation blind.

In the field of classical searchable encryption (SE), most schemes are either based on public key (RSA) \cite{Boneh04}, or based on symmetric key \cite{Song00,Curtmola06}. As we know, RSA has been theoretically broken by Shor algorithm \cite{Shor97} in polynomial time, while some symmetric cryptosystems, such as CBC-MAC, GMAC, GCM, etc., also have been recently broken by using quantum period finding \cite{Kaplan16,Dong19}. Therefore, how to use quantum technology to implement SE becomes an interesting work worth studying, which motivates us to study searchable encryption using blind quantum computation. Although some circuit-based BQC schemes \cite{Arrighi06,Tan17,Fisher14,Broadbent15,Zhang18,Liu18} have been continuously proposed in recent years, they only consider the single-client model, which cannot meet the requirements of multi-client accessing or searching data in the cloud environment. Besides, almost all of these schemes focus on guaranteeing the blindness of the data, while ignoring the blindness of the computation. The second motivation of our work is to implement the multi-client access mode as well as guarantee the blindness of computation.

This work designs a multi-client FBQC model, and utilizes it to propose a quantum SE scheme in cloud environment, but it maybe need some improvements or extensions in a practical one. In our scheme, the trusted key center hosts all the keys, so he becomes the cornerstone of the security and would also be the target of attacks (including quantum attacks). How to guarantee his security, i.e., to protect him from various attacks, will become an aspect to be explored. Second, how to prevent illegal user access in the multi-party FBQC model is not considered in this article, maybe quantum identity authentication (QIA) \cite{Li18} is a feasible one of the candidate solutions.

\section*{Acknowledgment}
The authors would like to express heartfelt gratitude to the anonymous reviewers and editor for their comments that improved the quality of this paper. And the support of all the members of the quantum research group of NUIST is especially acknowledged, their professional discussions and advice have helped us a lot.


\begin{thebibliography}{00}
\bibitem{Wang19}
F. Wang, J. Xu, X. Wang, S. G. Cui, "Joint offloading and computing optimization in wireless powered mobile-edge computing systems," IEEE Transactions on Wireless Communications, vol. 17, no. 3, pp. 1784-1797, 2018.

\bibitem{Xu19Li}
X. Xu, Y. Li, T. Huang, Y. Xue, K. Peng, L. Qi, W. Dou, "An energy-aware computation offloading method for smart edge computing in wireless metropolitan area networks," Journal of Network and Computer Applications, vol. 133, pp. 75-85, 2019.

\bibitem{Li19}
X. Li, G. Xu, X. Zheng, K. Liang, E. Panaousis, T. Li, W. Wang, C. Shen, "Using sparse representation to detect anomalies in complex WSNs," ACM Trans. Intell. Syst. Technol., vol. 10, no. 6, pp. 1-18, 2019.

\bibitem{Ota18}
K. Ota, M. Dong, J. Gui, A. Liu, "QUOIN: Incentive mechanisms for crowd sensing networks," IEEE Network, vol. 32, no. 2, pp. 114-119, 2018.

\bibitem{Qi16}
L. Qi, X. Xu, W. Dou, J. Yu, Z. Zhou, X. Zhang, "Time-aware IoE service recommendation on sparse data," Mobile Information Systems, vol. 2016, Article ID 4397061, 2016.

\bibitem{Xu19Liu}
X. Xu, Q. Liu, Y. Luo, K. Peng, X. Zhang, S. Meng, L. Qi, "A computation offloading method over big data for IoT-enabled cloud-edge computing," Future Generation Computer Systems, vol. 95, pp. 522-533, 2019.

\bibitem{Ren18}
H. Ren, H. Li, Y. Dai, K. Yang, X. Lin, "Querying in internet of things with privacy preserving: challenges, solutions and opportunities," IEEE Network, vol. 32, no. 6, pp. 144-151, 2018.

\bibitem{Xu19DAD}
G. Xu, Y. Zhang, L. Jiao, E. Panaousis, K. Liang, H. Wang, X. Li, "DT-CP: A daouble-TTPs based contract-signing protocol with lower computational cost," IEEE Access, 2019. DOI: 10.1109/ACCESS.2019.2952213


\bibitem{Xu18He}
X. Xu, C. He, Z. Xu, L. Qi, S. Wan, M. Z. A. Bhuiyan, "Joint optimization of offloading utility and privacy for edge computing enabled IoT," IEEE Internet of Things Journal, DOI: 10.1109/JIOT.2019.2944007

\bibitem{Qi19Chen}
L. Qi, Y. Chen, Y. Yuan, S. Fu, X. Zhang, X. Xu, "A QoS-aware virtual machine scheduling method for energy conservation in cloud-based cyber-physical systems", World Wide Web, 2019. DOI: 10.1007/s11280-019-00684-y

\bibitem{Xu19Zhang}
X. Xu, X. Zhang, H. Gao, Y. Xue, L. Qi, W. Dou, "BeCome: Blockchain-enabled computation offloading for IoT in mobile edge computing," IEEE Transactions on Industrial Informatics. 2019. DOI: 10.1109/TII.2019.2936869

\bibitem{Qi19Wang}
L. Qi, R. Wang, C. Hu, S. Li, Q. He, X. Xu, "Time-aware distributed service recommendation with privacy-preservation," Information Sciences, vol. 480, pp. 354-364, 2019.

\bibitem{Xu19SSP}
G. Xu, W. Wang, L. Jiao, X. Li, K. Liang, X. Zheng, W. Lian, H. Xian, H. Gao, "SoProtector: Safeguard privacy for native SO files in evolving mobile IoT applications," IEEE Internet of Things Journal, 2019. DOI: 10.1109/JIOT.2019.2944006


\bibitem{Gong18}
W. Gong, L. Qi, Y. Xu, "Privacy-aware multidimensional mobile service quality prediction and recommendation in distributed fog environment," Wireless Communications and Mobile Computing, vol. 2018, Article ID 3075849, 2018.

\bibitem{Qi18}
L. Qi, X. Zhang, W. Dou, C. Hu, C. Yang, J. Chen, "A two-stage locality-sensitive hashing based approach for privacy-preserving mobile service recommendation in cross-platform edge environment," Future Generation Computer Systems, vol. 88, pp. 636-643, 2018.

\bibitem{Wang19A}
H. Wang, S. Ma, H. Dai, "A rhombic dodecahedron topology for human-centric banking big data," IEEE Transactions on Computational Social Systems, vol. 6, no. 5, pp. 1095-1105, 2019.

\bibitem{Fu17}
F. Su, "A survey on big data analytics technologies," in Proc. Int. Conf. 5G Future Wireless Netw. Cham, Switzerland: Springer, 2017, pp. 359¨C370.

\bibitem{Wang19B}
H. Wang, S. Ma, H. N. Dai, M. Imran, T. Wang, "Blockchain-based data privacy management with nudge theory in open banking," Future Generation Computer Systems, 2019. DOI: 10.1016/j.future.2019.09.010

\bibitem{Bosch14}
C. B\"{o}sch, P. Hartel, W. Jonker, A. Peter, "A survey of provably secure searchable encryption," ACM Comput. Surv., vol. 47, no. 2, pp. 1801-1851, 2014.

\bibitem{Song00}
D. X. Song, D. Wagner, A. Perrig, "Practical techniques for searches on encrypted data," In: Proceeding 2000 IEEE Symposium on Security and Privacy. S\&P 2000, Berkeley, CA, USA, 2000, pp. 44-55.

\bibitem{Goh03}
E.-J. Goh, "Secure indexes," IACR Cryptology ePrint Archive, vol. 2003, pp. 216, 2003.

\bibitem{Curtmola06}
R. Curtmola, J. Garay, S. Kamara, R. Ostrovsky, "Searchable symmetric encryption: Improved definitions and efficient constructions," In: Proceedings of 13th  ACM Conference on Computer and Communications Security, USA, 2006, pp. 79-88.

\bibitem{Boneh04}
D. Boneh, D. C. Giovanni, R. Ostrovsky, G. Persiano, "Public key encryption with keyword search," In: Advances in Cryptology-EUROCRYPT 2004, Interlaken, Switzerland, 2004, pp. 506-522.

\bibitem{Xia16}
Z. Xia, X. Wang, L. Zhang, Z. Qin, X. Sun, K. Ren, "A privacy-preserving and copy-deterrence content-based image retrieval scheme in cloud computing," IEEE Transactions on Information Forensics and Security, vol. 11, no. 11, pp. 2594-2608, 2016.

\bibitem{Xu19E}
G. Xu, H. Li, Y. Dai, K. Yang and X. Lin, "Enabling efficient and geometric range query with access control over encrypted spatial data," IEEE Transactions on Information Forensics and Security, vol. 14, no. 4, pp. 870-885, 2019.

\bibitem{Fu16}
Z. Fu, K. Ren, J. Shu, X. Sun, F. Huang, "Enabling personalized search over encrypted outsourced data with efficiency improvement," IEEE Transactions on Parallel and Distributed Systems, vol. 27, no. 9, pp. 2546-2559, 2016.

\bibitem{Chen19}
L. Chen, W.-K. Lee, C.-C. Chang, K.-K. R. Choo, N. Zhang, "Blockchain based searchable encryption for electronic health record sharing," Future Generation Computer Systems, vol. 95, pp. 420-429, 2019.

\bibitem{Wang19Guo}
H. Wang, C. Guo, S. Cheng, "LoC-A new financial loan management system based on smart contracts," Future Generation Computer Systems, vol. 100, pp. 648-655, 2019.

\bibitem{Cao14}
N. Cao, C. Wang, M. Li, K. Ren, W. Lou, "Privacy-preserving multi-keyword ranked search over encrypted cloud data," IEEE Transactions on Parallel and Distributed Systems, vol. 25, no. 1, pp. 222-233, 2014.

\bibitem{Shor97}
P.W. Shor, "Polynomial-time algorithms for prime factorization and discrete logarithms on a quantum computer," SIAM J. Comput., vol. 26, no. 5, pp. 1484-1509, 1997.

\bibitem{Kaplan16}
M. Kaplan, G. Leurent, A. Leverrier, M. Naya-Plasencia, "Breaking symmetric cryptosystems using quantum period finding," In: CRYPTO 2016, LNCS, vol. 9815, 2016, pp. 207-237.

\bibitem{Dong19}
X. Dong, Z. Li, X. Wang, "Quantum cryptanalysis on some generalized Feistel schemes," Science China Information Sciences, vol. 62, no. 2, pp. 22501, 2019.

\bibitem{Huang17}
W. Huang, Q. Su, B. Liu, Y. H. He, F. Fan, and B. J. Xu, "Efficient multiparty quantum key agreement with collective detection," Scientific Reports, vol. 7, no. 1, pp. 15264, 2017.

\bibitem{Liu18Xu}
W. J. Liu, Y. Xu, C. N. Yang, P. P. Gao, and W. B. Yu, "An efficient and secure arbitrary n-party quantum key agreement protocol using bell states," International Journal of Theoretical Physics, vol. 57, no. 1, pp. 195-207, 2018.

\bibitem{Wu18}
F. Wu, X. Zhang, W. Yao, Z. Zheng, L. Xiang, and W. Li, "An advanced quantum-resistant signature scheme for cloud based on eisenstein ring," Computers, Materials \& Continua, vol. 56, no. 1, pp. 19-34, 2018.

\bibitem{Qu18Zhu} Z. Qu, T. Zhu, J. Wang, and X. Wang, "A novel quantum stegonagraphy based on brown states," Computers, Materials \& Continua, vol. 56, no. 1, pp. 47-59, 2018.

\bibitem{Liu18Gao}
W. J. Liu, P. P. Gao, W. B. Yu, Z. G. Qu, and C. N. Yang, "Quantum relief algorithm," Quantum Information Processing, vol. 17, no. 10, pp. 280, 2018.

\bibitem{Liu19Gao}
W. J. Liu, P. P. Gao, Y. X. Wang, W. B. Yu, and M. J. Zhang, "A unitary weights based one-iteration quantum perceptron algorithm for non-ideal training sets," IEEE Access, vol. 7, no. 4, pp. 36854-36865, 2019.

\bibitem{Broadbent09}
A. Broadbent, J. Fitzsimons, E. Kashefi, "Universal blind quantum computation," In: 50th Annual IEEE Symposium on Foundations of Computer Science, Atlanta, GA, 2009, pp. 517--526.

\bibitem{Kong16}
X. Kong, Q. Li, C. Wu, F. Yu, J. He, Z. Sun, "Multiple-server flexible blind quantum computation in networks," International Journal of Theoretical Physics, vol. 55, no. 6, pp. 3001-3007, 2016.

\bibitem{Kashefi17}
E. Kashefi, A. Pappa, "Multiparty delegated quantum computing," Cryptography, vol. 1, no. 2, pp. 12, 2017.

\bibitem{Arrighi06}
P. Arrighi, L. Salvail, "Blind quantum computation," International Journal of Quantum Information, vol. 4, no. 5, pp. 883-898, 2006.

\bibitem{Tan17}
X. Tan, X. Zhou, "Universal half-blind quantum computation," Annals of Telecommunications, vol. 72, no. 9, pp. 589-595, 2017.

\bibitem{Fisher14}
K. A. G. Fisher, A. Broadbent, L. K. Shalm, Z. Yan, J. Lavoie, R. Prevedel, T. Jennewein, K. J. Resch, "Quantum computing on encrypted data," Nature Communications, vol. 5, pp. 3074, 2014.

\bibitem{Broadbent15}
A. Broadbent, "Delegating private quantum computations," Canadian Journal of Physics, vol. 93, no. 9, pp. 941-946, 2015.

\bibitem{Zhang18}
X. Zhang, J. Weng, X. Li, W. Luo, X. Tan, T. Song, "Single-server blind quantum computation with quantum circuit model," Quantum Information Processing, vol. 17, no.6, pp. 134, 2018.

\bibitem{Liu18}
W. Liu, Z. Chen, J. Liu, Z. Su, L. Chi, "Full-blind delegating private quantum computation," Computers, Materials \& Continua, vol. 56, no. 2, pp. 211-223, 2018.

\bibitem{Sheng18}
Y. B. Sheng, L. Zhou, "Blind quantum computation with a noise channel" Phys. Rev. A, vol. 98, no. 5, pp. 052343, 2018.

\bibitem{Fitzsimons17}
J. F. Fitzsimons, E. Kashefi, "Unconditionally verifiable blind quantum computation", Phys. Rev. A, vol. 96, no. 1, pp. 012303, 2017.

\bibitem{Li18}
Q. Li, Z. Li, W. H. Chan, S. Zhang, C. Liu, "Blind quantum computation with identity authentication," Physics Letters A, vol. 382, no. 14, pp. 938-941, 2018.

\bibitem{Nielsen02}
M. A. Nielsen, and I. Chuang, Quantum Computation and Quantum Information. New York, NY, USA: Cambridge University Press, 2002, pp. 20-30.

\bibitem{Bennett84}
C. H. Bennett, G. Brassard, "Quantum cryptography: public key distribution and coin tossing," In International conference on Computers, Systems \& Signal Processing, Springer, Bangalore, India, 1984, pp. 175--179.

\end{thebibliography}
\end{document}